\documentclass{aa}
\usepackage{txfonts}
\usepackage{epsfig}
\usepackage{subfigure}

\usepackage{times}
\usepackage{natbib}
\bibpunct{(}{)}{;}{a}{}{,}

\begin{document}

\title{The origin of the iron lines in NGC~7213}
\author{S. Bianchi, G. Matt, I. Balestra, G.C. Perola}

\offprints{Stefano Bianchi\\ \email{bianchi@fis.uniroma3.it}}

\institute{Dipartimento di Fisica, Universit\`a degli Studi Roma Tre, Italy}

\date{Received / Accepted}

\authorrunning{S. Bianchi et al.}

\abstract{The analysis of a simultaneous XMM-\textit{Newton}/BeppoSAX plus three previous BeppoSAX observations revealed that NGC~7213 is a rather peculiar Seyfert 1. No significant Compton reflection component was observed, while an iron line complex, best explained in terms of three narrow lines, is clearly apparent in the data. Due to the absence of the reflection component, the neutral iron line is likely not produced in a Compton-thick material, but current data do not allow to choose between a Compton-thin torus and the BLR. The two ionized iron lines from  \ion{Fe}{xxv} and \ion{Fe}{xxvi} may be produced in a photoionized gas with a column density of a few 10$^{23}$ cm$^{-2}$, in analogy with the case of NGC~5506.

}

\maketitle

\section{Introduction}

Iron K$\alpha$ emission lines are common features in Seyfert 1s. The profile of the line provides fundamental information on their origin. If it is produced in the innermost regions of the accretion disk, kinematic and relativistic effects contribute to forge a broad, asymmetric, double-peaked profile \citep[see][for a review]{fab00}. A systematic study of the iron line profiles in more than 20 Seyfert 1s observed by \textit{ASCA} suggested that most of them are likely to have broad lines \citep{nan97,rey97}.

Rather surprisingly, this scenario is not being confirmed by XMM-\textit{Newton} and \textit{Chandra}. 
Relativistically broadened lines are detected in only a few sources \citep{nan99,wilm01,turner02,fab02},
and often the profiles and/or time behaviour are not what expected from simple models.
The reason for this lack of observed relativistic lines may reside in the physical properties, e.g. the ionization state, of the accretion disk itself.

On the other hand, a narrow component of the iron line is almost invariably present in $Chandra$ high energy gratings 
and XMM CCD spectra \citep[e.g.][]{kaspi01,yaq01,reev01,pounds01,gond01,Matt01}. The line, typically unresolved at the energy resolution of the CCDs, must be produced far from the nucleus, either in the torus envisaged in the Unification model \citep{antonucci93} or in the Broad Line Region. In the former case, and if the torus is Compton-thick, a Compton reflection component should also be present, which can be best studied when simultaneous XMM/BeppoSAX observations are available, in order to exploit the unprecedented sensitivity of XMM at the iron line energy and the hard X-ray coverage of the BeppoSAX
PDS instrument. On the other hand, if the line is produced in the BLR, a much fainter Compton reflection component is expected, and the intrinsic width of the iron line should be the same as that of the optical broad lines. 

The nuclear activity of NGC~7213 (z=0.005977) was discovered by the \textit{HEAO} A-2 satellite \citep{marsh79}. It was then included in the \citet{pic82} sample and classified as a Seyfert 1 because of the presence of broad optical emission lines and a strong blue continuum \citep{ph79,fh84}. In this Letter, we will discuss the X-ray spectrum of the source, analysing a simultaneous XMM-\textit{Newton}/BeppoSAX observation  and three further BeppoSAX observations, yet unpublished.\\

We will assume H$_0=70$ km/s/Mpc throughout the paper.

\section{\label{data}Observations and data reduction}

\begin{table}
\caption{\label{log}The log of all analysed observations and exposure times.}
\begin{center}
\begin{tabular}{llll}

\textbf{Date} & \textbf{Mission} & \textbf{Instrument} & \textbf{T$_\mathbf{exp}$ (ks)}  \\ 
\hline  
05/30/1999 & BeppoSAX & \textsc{mecs} & 47\\ 
&  & \textsc{pds} & 56\\ 
\hline  
11/20/1999 & BeppoSAX & \textsc{mecs} & 15\\ 
&  & \textsc{pds} & 34\\ 
\hline  
05/03/2000 & BeppoSAX & \textsc{mecs} & 46\\ 
&  & \textsc{pds} & 31\\ 
\hline
05/27/2001 & BeppoSAX & \textsc{mecs} & 61\\ 
&  & \textsc{pds} & 38\\
\hline 
05/29/2001 & XMM-\textit{Newton} & \textsc{epic-pn} & 30\\ 

\end{tabular} 
\end{center}

\end{table}

\subsection{XMM-Newton}

NGC~7213 was observed by XMM-\textit{Newton} on May 29 2001 (see Table \ref{log}) with the imaging CCD cameras, the EPIC-pn \citep{struder01} and the EPIC-MOS \citep{turner01}, adopting the Medium filter and operating in Prime Small Window and Prime Partial W2 respectively. The selected subarray allows a pileup free spectrum for the pn, since the observed count rate is much lower than the maximum defined for a 1\% pileup (see Table 3 of the XMM-Newton Users' Handbook). On the other hand, the MOS count rates are only slightly lower than the maximum value: a more careful inspection of the data with the \textsc{SAS} tool \textsc{epatplot}, together with a comparison with the pn spectra, suggests that the MOS data are probably affected by pileup. Therefore, we will not use the MOS data in this Letter. The data were reduced with \textsc{SAS} 5.4.1. X-ray events corresponding to pattern 0-4 were used for the pn and an extraction radius of 40$\arcsec$ was chosen for spectra and lightcurves. Spectra were analysed with \textsc{Xspec} 11.2.0.

\subsection{BeppoSAX}

BeppoSAX observed the source four times, from 1999 to 2001 (see Table \ref{log}). The last observation was performed between May 27 and May 30 2001, thus being simultaneous to the XMM-\textit{Newton} one. Data were downloaded from the ASDC archive\footnote{http://www.asdc.asi.it/} and analysed with \textsc{Xspec} 11.2.0. Since the softer part of the spectrum is beyond the scope of this work, LECS data will not be treated in this Letter. A normalization factor of 0.86 was adopted between the PDS and the MECS \citep{fiore99}, appropriate for PDS spectra extracted with fixed rise time threshold. When dealing with the fit of the simultaneous XMM-\textit{Newton}/BeppoSAX observation, we adopted a normalization factor of 0.888 between the EPIC-pn and the PDS, which was obtained normalizing the pn and the MECS spectra in a combined fit.\\

In the following, errors are at the 90\% confidence level for one interesting parameter ($\Delta \chi^2 =2.71$), where not otherwise stated.

\section{Data analysis}

\subsection{Flux and spectral variability}

The source showed a long-term variation in the 2-10 keV flux from 2 to 4$\times10^{-11}$ erg cm$^{-2}$ s$^{-1}$ (corresponding to a luminosity ranging from 1.5 to 3$\times10^{42}$ erg s$^{-1}$) when comparing the BeppoSAX observations. Short-term variations always less than 5\% around the mean value were also observed within single BeppoSAX observations. No significant differences in the continuum slope, that could be associated with the flux variability, are apparent in the data.

\subsection{Spectral analysis}

We first tried to fit the pn/PDS simultaneous observation in the energy band 0.5-220 keV, with a simple power law with an exponential cut off, absorbed by the Galactic column density $2.04\times10^{20}$ cm$^{-2}$, as found with the \textsc{Heasarc} nH tool\footnote{http://heasarc.gsfc.nasa.gov/cgi-bin/Tools/w3nh/w3nh.pl}. The result is an unacceptable value of the $\chi^2$ (1500/261 d.o.f.). A soft excess and complex absorption features are present at energies below $\simeq2.5$ keV, together with residuals around the iron line energy. On the other hand, no significant excess is left at higher energies. In this Letter, we will focus on the iron line properties and the Compton reflection component: therefore, we will ignore the spectrum below $2.5$ keV in the following analysis, where not otherwise stated.

\begin{figure}
\centerline{\epsfig{figure=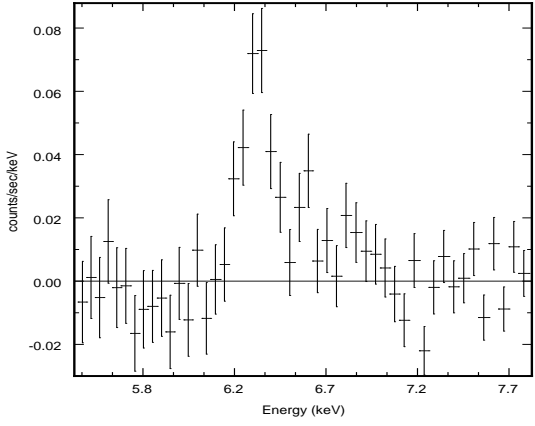, width=7.cm, height=5.5cm}}
\caption{\label{ironres}EPIC pn residuals around the iron line when no gaussian lines are added to the model (see text for details). The energies are not corrected for the redshift of the source.}
\end{figure}

The residuals around the iron line clearly suggest that it is complex (Fig. \ref{ironres}). If fitted with a single gaussian line, the centroid energy is $6.39^{+0.03}_{-0.01}$ keV and the intrinsic width $\sigma$ is $<90$ eV. The inclusion of a second gaussian line, with $E=6.68^{+0.06}_{-0.09}$ keV and $\sigma<250$ eV, improves the fit at the 99\% confidence level, according to F-test. A third narrow line ($6.94^{+0.05}_{-0.10}$) may also be present at the 96\% confidence level. The fit is now perfectly acceptable, with three narrow ($\sigma$ fixed to 1 eV) gaussian lines (173/178 d.o.f.). 

The presence of a cold iron line, if produced by Compton-thick matter, should come along with a significant Compton reflection component. Therefore, we decided to adopt the  \textsc{pexrav} model \citep{mz95}, which includes this component in terms of the solid angle R=$\Omega/2\pi$ subtended by the reflecting material. The fit does not improve at all (173/177 d.o.f.), and the Compton reflection component is completely undetected (see Table \ref{pnfit}). We further tried an ionized reflector model \citep{brf01}, to reproduce both the weak Compton hump and the ionized component of the iron line: the resulting $\chi^2$ is significantly worse than for the previous model (204/179 d.o.f.).

To verify if the lack of detection of the Compton reflection is affected by excluding the softer part of the data, we extended the best fit model to the broad band spectrum (0.5-220 keV). As already mentioned, residuals in excess to the model are clearly present below 2.5 keV. After allowing for a new minimization of the $\chi^2$, the fit is still unacceptable ($\chi^2$=539/256 d.o.f.), and absorption features are left in the softer part of the spectrum. The addition of a warm absorber (model \textsc{absori} in \textsc{Xspec}) improves marginally the fit and only an upper limit on the column density of the gas is found. On the other hand, if the residuals below 2.5 keV are treated like a genuine soft excess, including a black body to the original model, a good $\chi^2$ (287/252 d.o.f.) is achieved. In this case, the other spectral parameters, notably $\Gamma$ and R, remain unaltered with respect to the fit reported in Table \ref{pnfit}.
 
\begin{figure}
\centerline{\epsfig{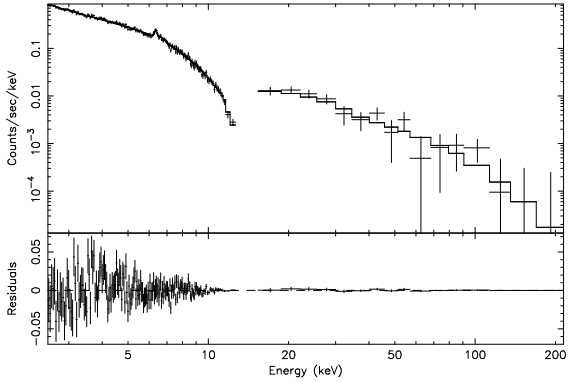}}
\caption{\label{pnspectrum}Best fit model and residuals for the pn/PDS spectrum (see text for details).}
\end{figure}

\begin{table}
\caption{\label{pnfit}Best fit parameters for the pn/PDS simultaneous observation. Th absorbing column density was fixed to that of our Galaxy, the intrinsic width of the lines was kept at the value of 1 eV because they are unresolved and a value of $i=30\degr$ was adopted for the unconstrained inclination angle (see text for details).}

\begin{center}
\begin{tabular}{ll}
\hline
$\mathbf\Gamma$ & $1.69\pm0.04$ \\ 
\textbf{E}$_\mathbf c$ \textbf{(keV)} & $95^{+75}_{-45}$ \\ 
\textbf{R} & $<0.20$ \\ 
\textbf{E}$_\mathbf1$ \textbf{(keV)} & $6.39\pm0.01$ \\ 
\textbf{EW}$_\mathbf1$ \textbf{(eV)}& $81^{+16}_{-12}$ \\ 
\textbf{F}$_\mathbf1$ \textbf{(10$^\mathbf{-5}$ ph cm$^\mathbf2$ s$^\mathbf{-1}$)} & 1.9$^{+0.4}_{-0.3}$\\
\textbf{E}$_\mathbf2$ \textbf{(keV)} & $6.65^{+0.04}_{-0.06}$\\
\textbf{EW}$_\mathbf2$ \textbf{(eV)}& $25^{+10}_{-13}$ \\ 
\textbf{F}$_\mathbf2$ \textbf{(10$^\mathbf{-5}$ ph cm$^\mathbf2$ s$^\mathbf{-1}$)} & 0.6$\pm0.3$\\
\textbf{E}$_\mathbf3$ \textbf{(keV)} & $6.94^{+0.05}_{-0.10}$ \\ 
\textbf{EW}$_\mathbf3$ \textbf{(eV)}& $22\pm14$ \\ 
\textbf{F}$_\mathbf3$ \textbf{(10$^\mathbf{-5}$ ph cm$^\mathbf2$ s$^\mathbf{-1}$)} & $0.5\pm0.3$\\
\textbf{F}$\mathrm{_{\mathbf {2-10\,keV}}}$ \textbf{(10$^\mathbf{-11}$ cgs)} & 2.2 \\
$\mathbf{\chi^2}$\textbf{/dof}  & 173/177 \\
\hline
\end{tabular} 
\end{center}

\end{table}

\begin{figure}
\centerline{\epsfig{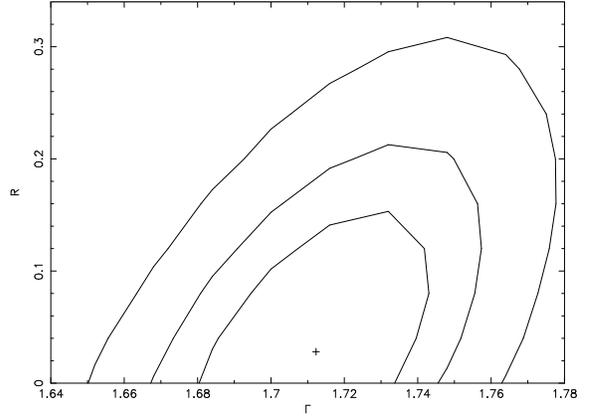}}
\caption{\label{gammar}$\Gamma$-R contour plot for the combined BeppoSAX observations. The curves refer to $\Delta\chi^{2}$=2.30, 4.61, 9.21, corresponding to confidence levels of 68, 90 and 99\% for two interesting parameters.}
\end{figure}

To look for the presence of the reflection component in the past history of the source, we decided to analyse all the available BeppoSAX observations (see Table \ref{log}). We adopted the best fit model used for the simultaneous observation, again ignoring data below $2.5$ keV. There is no significant detection of Compton reflection component in any of the observations, the value of R being always consistent with zero. Since no variation in the continuum slope is observed between the observations, we performed a combined fit of all of them, with all the parameters linked but the normalization factors needed to account for the different flux levels. In Fig. \ref{gammar} we show the $\Gamma$-R contour plots for this combined fit: the 90\% (for two parameters of interest) upper limit on R is $\simeq0.2$. This result is unique among bright Seyfert 1s observed by BeppoSAX \citep[e.g.][]{per02}. A similar case was claimed by \citet{weav01}: a very tight upper limit on the Compton reflection component in MCG-2-58-22 was found, based on \textit{ASCA} and \textit{RXTE} data, together with a significant iron line. However, results from a BeppoSAX/XMM-\textit{Newton} simultaneous observation suggest a higher value of R \citep{bianchi03b}.

Finally, we checked for a possible evidence of variability on the 6.4 keV line, leaving free the line normalizations between the different BeppoSAX observations in the combined fit: no significant improvement in the $\chi^2$ is found.

\section{Discussion} 

\begin{figure}
\centerline{\epsfig{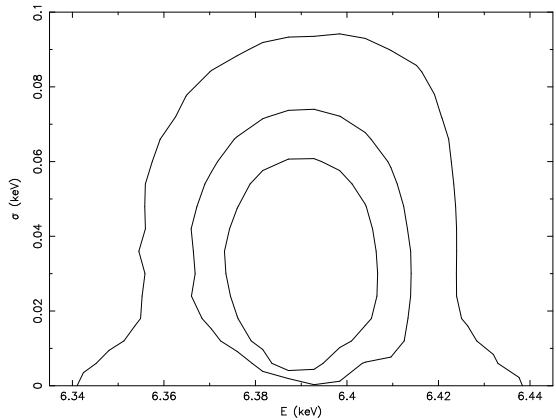}}
\caption{\label{linecontour}$\sigma$-E contour plot for the 6.4 keV iron line in the simultaneous BeppoSAX/XMM-\textit{Newton} observation, when the line width is left free to vary. The curves refer to $\Delta\chi^{2}$=2.30, 4.61, 9.21, corresponding to confidence levels of 68, 90 and 99\% for two interesting parameters.}
\end{figure}

The 6.4 keV line is clearly too narrow to be produced in the innermost region of the accretion disc, which would require a significant broadening of the feature (see Fig. \ref{linecontour}). Its origin must be placed in a material farther away from the BH, either Compton-thick, such as the 'torus', or Compton-thin, like the Broad Line Region. However, the notable absence of a Compton reflection component, as reported in the previous section, argues against an origin from Compton-thick material. In fact, a line with EW$\simeq80$ eV, as observed by XMM-\textit{Newton}, would require a reflection component with $R\simeq0.5$ \citep{matt91,gf91}. The only way to reconcile the data with the presence of a Compton-thick material would be assuming a torus with a small covering factor and an iron overabundance of about 3-5 \citep{mfr97}, at odds with typical values of about 1 found in Seyfert galaxies \citep{per02}.

Therefore, the line should originate in a Compton-thin material. \citet{mgm03} have shown that an iron line with EW$\simeq80$ eV can be produced by a face-on Compton-thin torus with N$_\mathrm{H}\simeq2\times10^{23}$ cm$^{-2}$. In this case, the associated reflection component would be much less prominent than for the Compton-thick material. A similar conclusion can be drawn if the line is produced in the BLR. To discriminate between the two possibilities, it is in principle possible to rely on the intrinsic width of the feature, which is significantly higher if the line is produced in the BLR, where the gas keplerian velocity is larger. Unfortunately, the 90\% (for two parameters of interest) EPIC-pn upper limit to the line FWHM is $8\,000$ km s$^{-1}$ (see Fig. \ref{linecontour}), which is similar to the broader component of the, indeed rather complex, H$\alpha$ line \citep{fh84}, thus providing no further information on the origin of the line. Only a \textit{Chandra} gratings observation could resolve the line width or put a tighter upper limit. No further informations can be derived from line variability. As reported in the previous section, the line flux is consistent with being constant among all the analysed observations, but the error bars are very large.

The observed centroid energies of the other two narrow lines indicate that they are produced by \ion{Fe}{xxv} and \ion{Fe}{xxvi}. The two ionized lines must therefore originate in a medium different from that of the neutral line, in analogy with the case of NGC~5506 \citep{Matt01,bianchi03}. \citet{bm02} have shown that a gas photoionized by a power law continuum can produce both lines, with EWs of a few tens of eV with respect to the total continuum: assuming a covering factor f=1 and a solar iron abundance, the required column density is around 10$^{23}$ cm$^{-2}$. A photoionized gas with such a column density would produce a continuum component with the same spectral shape of the primary emission and a flux of around 15$\%$. However, differently from NGC~5506, the primary continuum in NGC~7213 is not absorbed, preventing to disentangle the two components. A study of the variability of the ionized iron lines in response to the variation of the nuclear continuum would be useful in this case, but the BeppoSAX observations can provide only upper limits to the fluxes of these lines, making the investigation impossible.

\section{Conclusions}

NGC~7213 appears unique among Seyfert 1s observed by BeppoSAX, because it lacks a significant Compton reflection component in all of the four available observations performed by the Italian-Dutch satellite. The upper limit on R for the combined fit of the BeppoSAX observations is 0.2. Moreover, the EPIC pn spectrum, simultaneous with the last BeppoSAX observation, reveals the presence of an iron line complex, best explained in terms of a narrow neutral iron line with EW$\simeq$80 eV and two ionized lines from  \ion{Fe}{xxv} and \ion{Fe}{xxvi}, with EW of a few tens of eV. Because of the absence of a Compton reflection component, the neutral iron line is likely produced in the BLR or a Compton-thin torus, even if current data do not allow to make a choice between these two alternatives. The ionized lines may be produced in a photoionized gas with a column density of a few 10$^{23}$ cm$^{-2}$, but other solutions could not be excluded.

\acknowledgement

We would like to thank the referee (J.N.~Reeves) for his useful suggestions. SB, IB, GM and GCP acknowledge ASI for financial support.

\bibliographystyle{aa}
\bibliography{sbs}

\begin{thebibliography}{32}
\expandafter\ifx\csname natexlab\endcsname\relax\def\natexlab#1{#1}\fi

\bibitem[{{Antonucci}(1993)}]{antonucci93}
{Antonucci}, R. 1993, \araa, 31, 473

\bibitem[{{Ballantyne} {et~al.}(2001){Ballantyne}, {Ross}, \& {Fabian}}]{brf01}
{Ballantyne}, D.~R., {Ross}, R.~R., \& {Fabian}, A.~C. 2001, \mnras, 327, 10

\bibitem[{{Bianchi} {et~al.}(2003{\natexlab{a}}){Bianchi}, {Balestra}, {Matt},
  {Guainazzi}, \& {Perola}}]{bianchi03}
{Bianchi}, S., {Balestra}, I., {Matt}, G., {Guainazzi}, M., \& {Perola}, G.~C.
  2003{\natexlab{a}}, \aap, 402, 141

\bibitem[{{Bianchi} \& {Matt}(2002)}]{bm02}
{Bianchi}, S. \& {Matt}, G. 2002, \aap, 387, 76

\bibitem[{{Bianchi} {et~al.}(2003{\natexlab{b}}){Bianchi}, {Matt}, {Balestra},
  {Perola}, {}, \& {}}]{bianchi03b}
{Bianchi}, S., {Matt}, G., {Balestra}, I., {et~al.} 2003{\natexlab{b}}, in
  preparation

\bibitem[{{Fabian} {et~al.}(2000){Fabian}, {Iwasawa}, {Reynolds}, \&
  {Young}}]{fab00}
{Fabian}, A.~C., {Iwasawa}, K., {Reynolds}, C.~S., \& {Young}, A.~J. 2000,
  \pasp, 112, 1145

\bibitem[{{Fabian} {et~al.}(2002){Fabian}, {Vaughan}, {Nandra}, {Iwasawa},
  {Ballantyne}, {Lee}, {De Rosa}, {Turner}, \& {Young}}]{fab02}
{Fabian}, A.~C., {Vaughan}, S., {Nandra}, K., {et~al.} 2002, \mnras, 335, L1

\bibitem[{{Filippenko} \& {Halpern}(1984)}]{fh84}
{Filippenko}, A.~V. \& {Halpern}, J.~P. 1984, \apj, 285, 458

\bibitem[{{Fiore} {et~al.}(1999){Fiore}, {Guainazzi}, \& {Grandi}}]{fiore99}
{Fiore}, F., {Guainazzi}, M., \& {Grandi}, P. 1999, SDC report
  (http://www.asdc.asi.it/bepposax/software/index.html)

\bibitem[{{George} \& {Fabian}(1991)}]{gf91}
{George}, I.~M. \& {Fabian}, A.~C. 1991, \mnras, 249, 352

\bibitem[{{Gondoin} {et~al.}(2001){Gondoin}, {Barr}, {Lumb}, {Oosterbroek},
  {Orr}, \& {Parmar}}]{gond01}
{Gondoin}, P., {Barr}, P., {Lumb}, D., {et~al.} 2001, \aap, 378, 806

\bibitem[{{Kaspi} {et~al.}(2001){Kaspi}, {Brandt}, {Netzer}, {George},
  {Chartas}, {Behar}, {Sambruna}, {Garmire}, \& {Nousek}}]{kaspi01}
{Kaspi}, S., {Brandt}, W.~N., {Netzer}, H., {et~al.} 2001, \apj, 554, 216

\bibitem[{{Magdziarz} \& {Zdziarski}(1995)}]{mz95}
{Magdziarz}, P. \& {Zdziarski}, A.~A. 1995, \mnras, 273, 837

\bibitem[{{Marshall} {et~al.}(1979){Marshall}, {Boldt}, {Holt}, {Mushotzky},
  {Rothschild}, {Serlemitsos}, \& {Pravdo}}]{marsh79}
{Marshall}, F.~E., {Boldt}, E.~A., {Holt}, S.~S., {et~al.} 1979, \apjs, 40, 657

\bibitem[{{Matt} {et~al.}(1997){Matt}, {Fabian}, \& {Reynolds}}]{mfr97}
{Matt}, G., {Fabian}, A.~C., \& {Reynolds}, C.~S. 1997, \mnras, 289, 175

\bibitem[{{Matt} {et~al.}(2003){Matt}, {Guainazzi}, \& {Maiolino}}]{mgm03}
{Matt}, G., {Guainazzi}, M., \& {Maiolino}, R. 2003, MNRAS in press,
  astro-ph/0302328

\bibitem[{{Matt} {et~al.}(2001){Matt}, {Guainazzi}, {Perola}, {Fiore},
  {Nicastro}, {Cappi}, \& {Piro}}]{Matt01}
{Matt}, G., {Guainazzi}, M., {Perola}, G.~C., {et~al.} 2001, \aap, 377, L31

\bibitem[{{Matt} {et~al.}(1991){Matt}, {Perola}, \& {Piro}}]{matt91}
{Matt}, G., {Perola}, G.~C., \& {Piro}, L. 1991, \aap, 247, 25

\bibitem[{{Nandra} {et~al.}(1997){Nandra}, {George}, {Mushotzky}, {Turner}, \&
  {Yaqoob}}]{nan97}
{Nandra}, K., {George}, I.~M., {Mushotzky}, R.~F., {Turner}, T.~J., \&
  {Yaqoob}, T. 1997, \apj, 477, 602

\bibitem[{{Nandra} {et~al.}(1999){Nandra}, {George}, {Mushotzky}, {Turner}, \&
  {Yaqoob}}]{nan99}
---. 1999, \apjl, 523, L17

\bibitem[{{Perola} {et~al.}(2002){Perola}, {Matt}, {Cappi}, {Fiore},
  {Guainazzi}, {Maraschi}, {Petrucci}, \& {Piro}}]{per02}
{Perola}, G.~C., {Matt}, G., {Cappi}, M., {et~al.} 2002, \aap, 389, 802

\bibitem[{{Phillips}(1979)}]{ph79}
{Phillips}, M.~M. 1979, \apjl, 227, L121+

\bibitem[{{Piccinotti} {et~al.}(1982){Piccinotti}, {Mushotzky}, {Boldt},
  {Holt}, {Marshall}, {Serlemitsos}, \& {Shafer}}]{pic82}
{Piccinotti}, G., {Mushotzky}, R.~F., {Boldt}, E.~A., {et~al.} 1982, \apj, 253,
  485

\bibitem[{{Pounds} {et~al.}(2001){Pounds}, {Reeves}, {O'Brien}, {Page},
  {Turner}, \& {Nayakshin}}]{pounds01}
{Pounds}, K., {Reeves}, J., {O'Brien}, P., {et~al.} 2001, \apj, 559, 181

\bibitem[{{Reeves} {et~al.}(2001){Reeves}, {Turner}, {Pounds}, {O'Brien},
  {Boller}, {Ferrando}, {Kendziorra}, \& {Vercellone}}]{reev01}
{Reeves}, J.~N., {Turner}, M.~J.~L., {Pounds}, K.~A., {et~al.} 2001, \aap, 365,
  L134

\bibitem[{{Reynolds}(1997)}]{rey97}
{Reynolds}, C.~S. 1997, \mnras, 286, 513

\bibitem[{{Str{\" u}der} {et~al.}(2001){Str{\" u}der}, {Briel}, {Dennerl},
  {Hartmann}, {Kendziorra}, {Meidinger}, {Pfeffermann}, {Reppin}, {Aschenbach},
  {Bornemann}, {Br{\" a}uninger}, {Burkert}, {Elender}, {Freyberg}, {Haberl},
  {Hartner}, {Heuschmann}, {Hippmann}, {Kastelic}, {Kemmer}, {Kettenring},
  {Kink}, {Krause}, {M{\" u}ller}, {Oppitz}, {Pietsch}, {Popp}, {Predehl},
  {Read}, {Stephan}, {St{\" o}tter}, {Tr{\" u}mper}, {Holl}, {Kemmer},
  {Soltau}, {St{\" o}tter}, {Weber}, {Weichert}, {von Zanthier},
  {Carathanassis}, {Lutz}, {Richter}, {Solc}, {B{\" o}ttcher}, {Kuster},
  {Staubert}, {Abbey}, {Holland}, {Turner}, {Balasini}, {Bignami}, {La
  Palombara}, {Villa}, {Buttler}, {Gianini}, {Lain{\' e}}, {Lumb}, \&
  {Dhez}}]{struder01}
{Str{\" u}der}, L., {Briel}, U., {Dennerl}, K., {et~al.} 2001, \aap, 365, L18

\bibitem[{{Turner} {et~al.}(2001){Turner}, {Abbey}, {Arnaud}, {Balasini},
  {Barbera}, {Belsole}, {Bennie}, {Bernard}, {Bignami}, {Boer}, {Briel},
  {Butler}, {Cara}, {Chabaud}, {Cole}, {Collura}, {Conte}, {Cros}, {Denby},
  {Dhez}, {Di Coco}, {Dowson}, {Ferrando}, {Ghizzardi}, {Gianotti}, {Goodall},
  {Gretton}, {Griffiths}, {Hainaut}, {Hochedez}, {Holland}, {Jourdain},
  {Kendziorra}, {Lagostina}, {Laine}, {La Palombara}, {Lortholary}, {Lumb},
  {Marty}, {Molendi}, {Pigot}, {Poindron}, {Pounds}, {Reeves}, {Reppin},
  {Rothenflug}, {Salvetat}, {Sauvageot}, {Schmitt}, {Sembay}, {Short},
  {Spragg}, {Stephen}, {Str{\" u}der}, {Tiengo}, {Trifoglio}, {Tr{\" u}mper},
  {Vercellone}, {Vigroux}, {Villa}, {Ward}, {Whitehead}, \& {Zonca}}]{turner01}
{Turner}, M.~J.~L., {Abbey}, A., {Arnaud}, M., {et~al.} 2001, \aap, 365, L27

\bibitem[{{Turner} {et~al.}(2002){Turner}, {Mushotzky}, {Yaqoob}, {George},
  {Snowden}, {Netzer}, {Kraemer}, {Nandra}, \& {Chelouche}}]{turner02}
{Turner}, T.~J., {Mushotzky}, R.~F., {Yaqoob}, T., {et~al.} 2002, \apjl, 574,
  L123

\bibitem[{{Weaver}(2001)}]{weav01}
{Weaver}, K.~A. 2001, in AIP Conf. Proc. 599: X-ray Astronomy: Stellar
  Endpoints, AGN, and the Diffuse X-ray Background, 482--+

\bibitem[{{Wilms} {et~al.}(2001){Wilms}, {Reynolds}, {Begelman}, {Reeves},
  {Molendi}, {Staubert}, \& {Kendziorra}}]{wilm01}
{Wilms}, J., {Reynolds}, C.~S., {Begelman}, M.~C., {et~al.} 2001, \mnras, 328,
  L27

\bibitem[{{Yaqoob} {et~al.}(2001){Yaqoob}, {George}, {Nandra}, {Turner},
  {Serlemitsos}, \& {Mushotzky}}]{yaq01}
{Yaqoob}, T., {George}, I.~M., {Nandra}, K., {et~al.} 2001, \apj, 546, 759

\end{thebibliography}

\end{document}